\documentclass{aa}  
\usepackage{graphicx}
\usepackage{txfonts}
\usepackage[colorlinks=true,linkcolor=blue,citecolor=blue]{hyperref}
\usepackage{appendix}
\defcitealias{khetan2021new}{K21}

\begin{document}

   \title{The Hubble constant from two sibling Type Ia supernovae in the nearby galaxy NGC~4414: SN~1974G and SN~2021J}

  % \subtitle{SN2021J}

   \author{Eulalia Gallego-Cano
          \inst{1}
          \and
          Luca Izzo
          \inst{2}
          \and
          Carlos Dominguez-Tagle
          \inst{1}        
          \and
          Francisco Prada
          \inst{1}
          \and \\
          Enrique P\'erez
          \inst{1}
          \and 
          Nandita Khetan
          \inst{2}
          \and
          In Sung Jang
          \inst{3}
   }

   \institute{
       Instituto de Astrof\'isica de Andaluc\'ia (CSIC),
     Glorieta de la Astronom\'ia s/n, 18008 Granada, Spain; \email{lgc@iaa.es} 
     \and
     DARK, Niels Bohr Institute, University of Copenhagen, Jagtvej 128, 2200 Copenhagen, Denmark
     \and
     Department of Astronomy \& Astrophysics, University of Chicago, 5640 South Ellis Avenue, Chicago, IL 60637, USA
   }

   \date{Accepted XXX. Received YYY; in original form ZZZ}

% \abstract{}{}{}{}{} 
% 5 {} token are mandatory
 
  \abstract
   {Having two ``sibling'' Type Ia supernovae (SNe Ia) in the same galaxy offers additional advantages in reducing a variety of systematic errors involved in estimating the Hubble constant, $H_{0}$. NGC~4414 is a nearby galaxy included in the \textit{Hubble Space Telescope} Key Project to measure its distance using Cepheid variables. It hosts two sibling SNe Ia: SN~2021J and SN~1974G. This provides the opportunity to improve the precision of the previous estimate of $H_{0}$, which was based solely on SN~1974G. Here we present new optical $BVRI$ photometry obtained at the Observatorio de Sierra Nevada and complement it with {\it Swift} UVOT $UBV$ data, which cover the first 70 days of emission of SN~2021J. A first look at SN~2021J optical spectra obtained with the Gran Telescopio Canarias (GTC) reveals typical SN type Ia features. The main SN luminosity parameters for the two sibling SNe are obtained by using SNooPy, a light curve fitting code based on templates. Using a hierarchical bayesian approach, we build the Hubble diagram with a sample of 96 SNe~Ia obtained from the Combined Pantheon Sample in the redshift range $z = 0.02-0.075$, and calibrate the zero point with the two sibling type-Ia SNe in NGC~4414. We report a value of the Hubble constant $H_{0}$ $= 72.19 \pm 2.32$ (stat.) $\pm 3.42$ (syst.) km s$^{-1}$Mpc$^{-1}$. We expect a reduction of the systematic error after a new analysis of the Cepheids period-luminosity relation using the upcoming Gaia DR4 and additional Cepheids from the HST and JWST.
   } 

{}
  {}

  {}


  \titlerunning{The Hubble constant from two SNe~Ia sibling in the nearby galaxy NGC~4414}
  \authorrunning{E. Gallego-Cano et al.}

   \maketitle
%
%-------------------------------------------------------------------
\section{Introduction}

Type Ia supernovae (SNe~Ia) play an essential role as distance indicators for cosmological studies that allow us to explore far out into the Hubble flow. SNe~Ia have been standardized by applying various corrections to their absolute luminosity, decline-absolute magnitude and intrinsic color-reddening relationships \citep{phillips1993absolute,hamuy19931990,riess1996precise,tripp1997two}. Hence, SNe~Ia have been widely used to determine the Hubble constant ($H_{0}$), which gives an estimate of the expansion rate of the Universe \citep{hubble1929spiral}. This is a key parameter of the standard $\Lambda$CDM\footnote{$\Lambda$ cold dark matter model, where $\Lambda$ represents the cosmological constant.} cosmological model \citep{peebles2012seeing}. However, their estimated values at the two extremes of the Universe, in early times from the cosmic microwave background (CMB) \citep{2020A&A...641A...6P} and in the local Universe from classical methods \citep[e.g.][]{freedman2001final,riess20162}, show a discrepancy \citep[see][for a review]{verde2019tensions,riess2021comprehensive}. This tension may indicate new physics beyond the standard cosmological model. Therefore, it is mandatory to explore the possible systematics in the local values of $H_{0}$ and obtain more precise measurements. 

The nearby galaxy NGC~4414 is an isolated member of the Coma~I group \citep{braine1997anatomy} and is one of the galaxies included in the Hubble Space Telescope (HST) Key Project on the extragalactic distance scale using Cepheid variables stars \citep{freedman2001final}. This galaxy hosts the type-Ia supernova SN~1974G, which has been used in previous studies to measure $H_{0}$ \citep{schaefer1998peak,gibson2000hubble}, and the type IIb supernova SN~2013df \citep{2013CBET.3557....1C,2014AJ....147...37V}. Type~II SNe can also be considered as distance indicators, however, their study would need to be further improved to obtain accurate extragalactic distances \citep[e.g.][]{2020MNRAS.495.4860D,2022arXiv220308974D}. The recently discovered type-Ia supernova SN~2021J, by the Zwicky Transient Factory (ZTF, \citealp{Bellm2019}), hosted in the same galaxy brings us the opportunity to improve and reduce possible systematic errors in the estimate of the Hubble constant by using two “siblings” supernovae from a single nearby galaxy, such as those related to the host galaxy distance, the peculiar velocity, and other unaccounted properties of the host galaxy \citep[see][for improving the relative precision of SNe Ia as standard candles using siblings SN in nearby galaxies]{burns2020sn}. Moreover, studying siblings offers us a great opportunity to explore many essential aspects related to the cosmological utility of SNe~Ia, such as the correlations of SN light curves (LCs), and properties of their host galaxies, including the intrinsic dust extinction, and the validity of the standardization procedure \citep{scolnic2020supernova,biswas2022two}.

In this paper, we present BVRI photometry of SN~2021J in eight epochs obtained with the CCDT150 camera mounted on the 1.5-m telescope at the Observatorio de Sierra Nevada (OSN), and UBV photometry from {\it Swift}-UVOT. Moreover, we report multi-epoch spectra observed with the OSIRIS spectrograph on the Gran Telescopio Canarias (GTC) at La Palma. From the LC analysis of the 2 siblings SN~1974G and SN~2021J in NGC~4414, we determine the Hubble constant by means of a hierarchical bayesian approach using a SNe~Ia sample drawn from the Combined Pantheon Sample in the redshift range $z = 0.02-0.075$. In Sec.\ref{observations} we present our SN~2021J photometric and spectroscopic data, and describe the data used for SN~1974G. In Sec.\ref{analysis} we focus on the analysis of both SN light-curves, and derive the Hubble constant. Finally, the results are summarised and discussed in Sec.\ref{conclusions}.
%--------------------------------------------------------------------
\section{Observations}
\label{observations}
\subsection{Photometry}
We used the $B-$ and $V-$band photometry of SN~1974G presented in \citet{schaefer1998peak}, where the author collected existing data from the literature and improved the photometric calibration, building updated $B-$ and $V-$band light curves. We note that the existing data contains photographic magnitudes, but if we remove them from our analysis, our results are still consistent within the margins of error. Finally, we decided to consider all the original points despite large uncertainties because the SN peak is better covered (see more details in Appendix \ref{unc_SN1974G}).

SN~2021J was discovered by ZTF in the galaxy NGC~4414 as a new object of magnitude $g$(AB)
 = 17 mag \citep{2021TNSTR...1....1F}, and it was observed as part of the program \emph{SN2: SuperNovae from Sierra Nevada}\footnote{\url{http://sn2.iaa.es/}}. This program aims at building a statistically significant BVRI photometric sample of SNe~Ia in nearby Sloan Digital Sky Survey (SDSS) host galaxies to study how their luminosity properties relates with host environment, and thus, how the latter can affect their physical properties. The optical images for SN~2021J were obtained with the CCDT150 camera mounted on the 1.5-m telescope at the Observatorio de Sierra Nevada (OSN) located in Granada, Spain (see Fig.\,\ref{SN2021J_multifilter}). We used the Johnson-Cousins BVRI filters to obtain the SN photometry in eight epochs during an observing interval of 60 days from the first observation. Two epochs were observed before the SN peak brightness, which allows to sample the peak luminosity and build a reliable light curve (LC). 
 
 The data have been reduced by using a Python-based\footnote{Python Software Foundation. Python Language Reference, version 3.9. Available at \url{http://www.python.org}} pipeline developed by us to automatically reduce the data obtained for the SN2 observing program. Bias subtraction and flat fielding were implemented using the \textsc{ccdproc} and \textsc{astropy} Python packages \citep{matt_craig_2017_1069648}. Subsequently plate-solve calibration was applied to single images using astrometry.net libraries, which were finally stacked into a single calibrated image using \textsc{Swarp} \citep{2002ASPC..281..228B}. Finally, we performed a differential photometry of the SN by using field stars whose BVRI magnitudes have been measured from observations of the SA98 Landolt field. Magnitudes have been measured with PSF photometry on all the stellar sources in each image having a brightness larger than twice the standard deviation value computed on the image itself, and then using an integrated Gaussian pixel response function as our model for the magnitude measurement. To this aim, we made a wide use of the tools provided within the \texttt{photutils}\footnote{https://photutils.readthedocs.io/en/stable/index.html} package; in the specific we used the \texttt{BasicPSFPhotometry} class for PSF photometry of all sources. More details will be provided in a dedicated SN2 sample paper (Izzo et al. in preparation), and the pipeline codes will be available in the website of the project. The photometry of SN~2021J is presented in Table\,\ref{Tab:photometry}.

We also reduced the {\it Swift} data available for SN~2021J using the most updated \texttt{HEASOFT} version (v6.30) and calibration database (CALDB). After correcting sky coordinates with \texttt{uvotskycorr}, we combined all images for each single observation epoch and performed the photometry measurements using the  calibration presented in \citet{Breeveld2010} with \texttt{uvotsource}, using a $5''$ aperture centered on SN~2021J and an annulus background of  $8''$ radius. The results of our measurements are given in Table \ref{Tab:swift}.

\subsection{Spectroscopy}

\begin{figure} [ht!]
\includegraphics[width=\columnwidth]{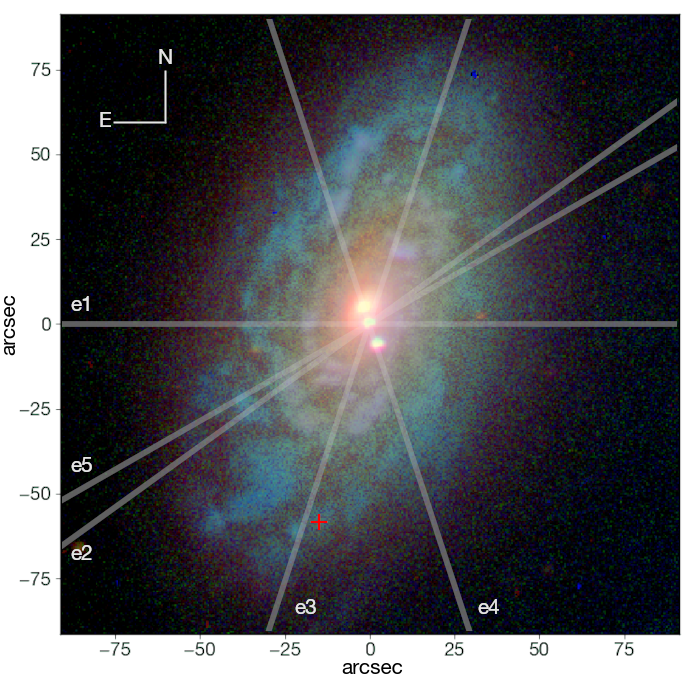}
\caption{Slit position for the five epochs (e1-e5) of observation over an OSN image of NGC~4414. All slits are centered on SN~2021J. The position of SN~1974G is marked with a red $+$ sign. The image shows a foreground star $5''$ SW apart of SN 2021J. }
  \label{slitorient}
\end{figure}

We also performed follow-up spectroscopic observations of SN~2021J during the Director's Discretionary Time program (ID: GTC11-20BDDT) using OSIRIS \citep{Jordi03} in the long-slit mode on the 10.4-m Gran Telescopio Canarias (GTC) at Observatorio del Roque de los Muchachos (ORM), La Palma, Spain. The observations span over five epochs, with spectra taken at the B-band peak brightness, 2 days before and 6, 22, and 163 days after the peak. Different high-resolution gratings were used to cover as much wavelength range as possible. Five slit position angles (PA) were adopted to achieve a more complete spatial coverage of NGC~4414 (see Fig.\,\ref{slitorient}). A slit width of $0.6^{\prime\prime}$ was used for all the observations, and the exposure times were selected according to the grating and object brightness at the epoch of observation. The basic parameters of these observations are given in Table\,\ref{Tab:spect}.

\begin{table}
\centering
\caption{Information on the OSIRIS/GTC spectroscopy of SN~2021J.}
\label{Tab:spect} 
\begin{tabular}{ccrcc}
\hline
\hline
Epoch & Date  	& PA            & Airmass 	& Wavelength range \\
      & (MJD)  	& ($^{\circ}$)  &  	           & (\AA) \\
\hline

1 &   59231.26	& 90		&    1.01	& 3950 - 10000		\\
2 &   59233.29	& -54		&    1.03	& 3950 - 10000		\\     
3 &   59239.16	& -18		&    1.07	& 3950 - 10000		\\       
4 &   59255.26	& 18		&    1.09	& 3950 - 10000		\\        
5 &   59395.91	& -60		&    1.26	& 3600 -  9300		\\       
\hline
\hline
\end{tabular}
 \end{table}
 
We performed the reduction of the spectra using IRAF standard tasks \citep{Tody86}, and GTCMOS pipeline \citep{GTCMOS16}. The full details of the data reduction and its analysis  will be presented in a subsequent publication. Here we present a brief summary. Bias and flat–field corrections were applied to the object and standard images. The 2D spectra were calibrated in wavelength using the HgAr+Ne+Xe lamps from the GTC instrument calibration module. The 1D spectra were extracted using an optimal extraction aperture, and the flux calibration was done using spectro-photometric standard stars, observed the same night and with the same gratings than SN~2021J.
The atmospheric extinction was corrected using the ORM extinction curve at the observed airmass. 
The dust extinction in the host galaxy of SN~2021J has not been corrected. 
The spectra over the different wavelength ranges were combined into a single spectrum for each epoch, and corrected from foreground Galactic reddening \citep{schlegel1998maps}.

\begin{figure} [ht!]
\includegraphics[width=0.95\columnwidth]{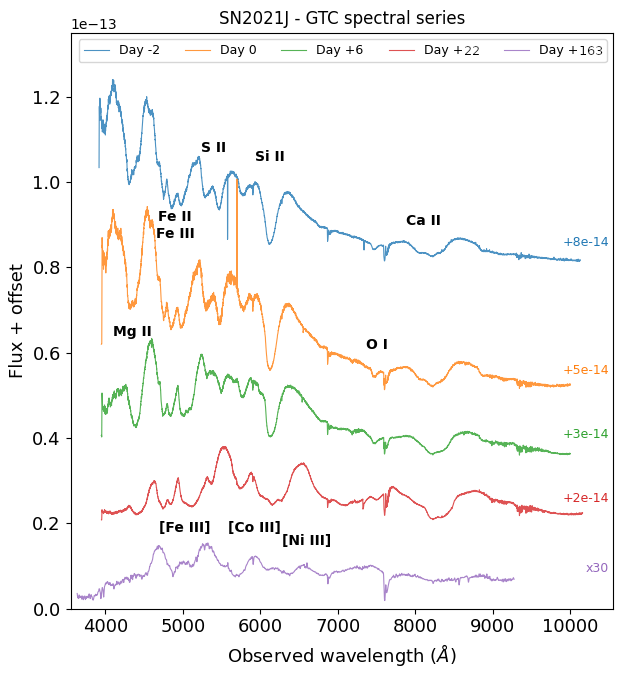}
\caption{Spectroscopic evolution of SN~2021J over the five epochs observed with OSIRIS/GTC. Each epoch (days from peak brightness) is plotted with a different color. Flux offset for each epoch are reported on the right end of each spectrum. The main line features are also reported, showing the type-Ia nature for SN~2021J.}
  \label{spectra}
  \centering
\end{figure}

The SN~2021J spectroscopic evolution over the five epochs is shown in Fig.\,\ref{spectra}. 
The strongest early-time features, seen prior to and near the peak brightness, are the Mg~II, S~II, Si~II, and Ca~II lines, which are typically observed in type-Ia SNe at similar epochs \citep{Gal-Yam2017}. 
The late nebular spectrum (day 163 after the peak) shows typical forbidden emission lines of [Fe~III] and [Co~III] that are generally observed at these late epochs of the SN explosion. The spectra are made publicly available on WISeREP\footnote{\url{https://www.wiserep.org}} \citep{2012PASP..124..668Y}. A detailed analysis of the spectral series of SN~2021J will be presented elsewhere.

%-----------------------------------------------------------------
\section{Analysis}
\label{analysis}
\subsection{The light curves of SN~1974G and SN~2021J}
The SNe light curves were fitted with the SNooPy (SuperNovae in object oriented Python) package \citep{burns2010carnegie}. The software generates uBVgriYJH light-curve templates to perform a LC fitting to determine the main luminosity parameters. We parameterized the SNe LCs using the \textit{decay-rate parameter} $\Delta m_{15}$ introduced by \cite{phillips1993absolute}, and defined as the difference in magnitude between the peak and 15 days after peak of SN emission in the $B$ band. We selected the \textit{max-model} implemented in SNooPy for the fits to determine in each filter the maximum magnitude and $\Delta m_{15}$. This model does not assume any intrinsic extinction of the SN, neither in the host galaxy nor in the intergalactic medium. The model only corrects for the Milky Way extinction using the Schlegel maps \citep{schlegel1998maps}, and it uses the spectral template of \cite{hsiao2007k} to compute K-corrections\footnote{For further information, see \url{https://csp.obs.carnegiescience.edu/data/snpy/documentation}}. We considered a value of $cz=991$ km s$^{-1}$ for the recessional velocity of NGC~4414 with respect to the CMB rest frame\footnote{From Cosmic flow model of our local Universe website. For more information see \url{https://cosmicflows.iap.fr/}} \citep{2015MNRAS.450..317C}.  

\begin{figure*} [ht!]
\begin{center}
\includegraphics[width=0.90\columnwidth]{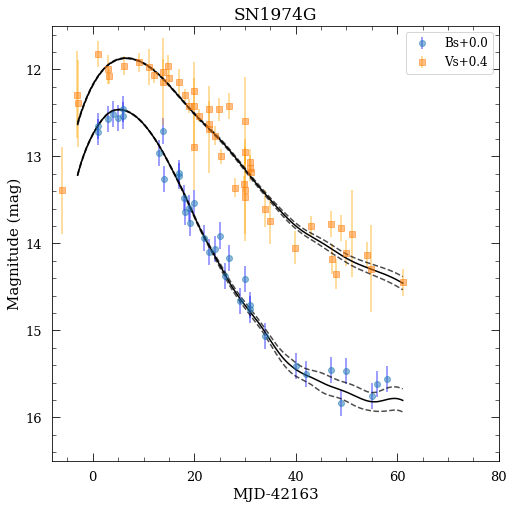}
\hspace{0.5cm}
\includegraphics[width=0.88\columnwidth]{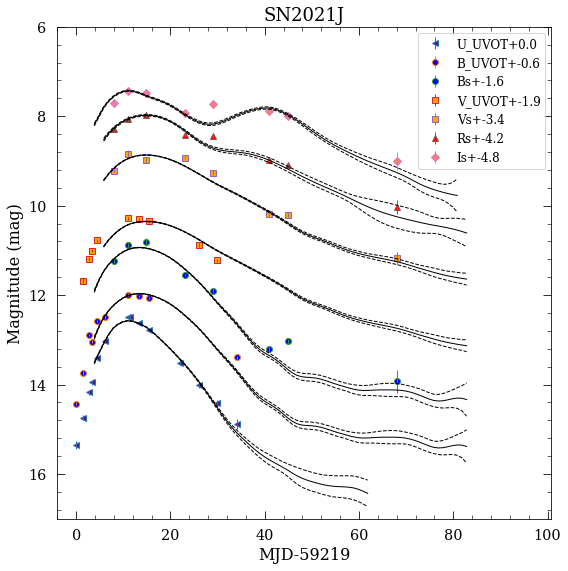}
\end{center}
\caption{{LC fits of the siblings using SNooPy. \it (Left)} SN~1974G LCs fits. The orange and blue points correspond to the $B$- and $V$-band data taken from \citet{schaefer1998peak}. The black lines indicate the SNooPy best-fits. Dashed lines are 1-$\sigma$ errors. {\it (Right)} SN~2021J LCs fits. The photometry data are obtained from the OSN and {\it Swift}. The legend indicates the color and shape of the symbols adopted for the different bands. The black lines indicate the SNooPy best-fits. Dashed lines are 1-$\sigma$ errors.}
  \label{SN2021J_multifilter}
\end{figure*}

Table\,\ref{Tab:snoopy} shows the LC parameters for both SNe. The first four rows correspond to the maximum brightness in the different filters. We note that SN~1974G has photometric data only in $B$ and $V$ bands. The fifth row indicates the MJD day of $B$-maximum, and the last row shows the decay-rate parameter. Figure\,\ref{SN2021J_multifilter} shows the LCs fits for SN~1974G and SN~2021J, respectively. The solid black lines are SNooPy fits for the different filters. Dashed lines in LC fits indicate the 1-$\sigma$ errors.

\begin{table}
\centering
\caption{SNooPy best-fit parameters for the LCs of SN~1974G and SN~2021J}
\label{Tab:snoopy} 
\begin{tabular}{lll}
\hline
\hline
Parameter & SN~1974G & SN~2021J \\
\hline
$U_{max}$ (mag) & - & $13.269 \pm 0.037$\\
$B_{max}$ (mag) & $12.417 \pm 0.043$ & $12.489 \pm 0.028$\\
$V_{max}$ (mag) & $12.252 \pm 0.035$ & $12.282 \pm 0.033$\\
$R_{max}$ (mag) & - & $12.196 \pm 0.075$\\
$I_{max}$ (mag) & - & $12.637 \pm 0.085$\\
$T_{max}$ (MJD) & $ 42168.65 \pm 0.50$ & $59233.00 \pm 0.16 $\\
$\bigtriangleup m_{15}$ (mag) & $1.236 \pm 0.074$ & $1.028 \pm 0.046$\\
\hline
\end{tabular}
 \end{table}
 
\subsection{The Hubble constant}
In order to measure the Hubble constant using these two sibling SNe in NGC~4414 as calibrators, we followed the procedure described in \citet[][hereafter K21]{khetan2021new}. 

First, we modeled the observed B magnitudes at their maximum based on the main luminosity parameters for type-Ia SNe, including two terms: one provided by \cite{phillips1993absolute} relating the peak luminosity of a SN~Ia and the LC shape, and the other provided by \cite{1998A&A...331..815T} including the color correction (see more details in \citetalias{khetan2021new}). To calibrate the model, here, we used SN~1974G and SN~2021J in the galaxy NGC~4414 with its distance modulus measured from Cepheid variables. By using $\bigtriangleup m_{15}$ to shape the LCs, instead of the \textit{color-stretch} used in \citetalias{khetan2021new}, the observed $B$ magnitude at peak is:
\begin{equation}
\label{eqn:tripp}
B_{max}=P_{0}+P_{1}(\bigtriangleup m_{15}-1.1)+R(B_{max}-V_{max})+\mu_{calib}
\end{equation}
where $P_{0}$ and $P_{1}$ relate the maximum magnitude to the decay-rate parameter $\Delta m_{15}$, $R$ includes the extinction correction that covers other possible causes of reddening, in addition to the galactic extinction that was already taken into account in the LC fitting. We note that our cosmological analysis does not require distinguishing between the different sources of extinction because our goal is not to study the properties of the dust in detail. ($B_{max}-V_{max}$) is the \textit{pseudo-color} defined by the difference between the maximum flux in $B$ and $V$ bands, and $\mu_{calib}$ is the distance modulus of the host galaxy. 

Several difficulties such as the galaxy compact size, its high surface brightness and a strong source crowding made the study of the Cepheids in NGC~4414 one of the most challenging in the HST Key Project. We adopted the distance modulus of $\mu_{NGC4414}=31.24 \pm 0.05$ (stat.) mag, as provided in the final results work of the Key Project \citep{freedman2001final}. They improved the value obtained by \cite{turner1998hubble} and \cite{gibson2000hubble} by including new corrections in the distance modulus to the Large Magellanic Cloud (LMC), new Wide Field and Planetary Camera 2 (WFPC2) photometric calibration, and improved corrections for metallicity and reddening on the observed period-luminosity (PL) relation of the Cepheids. We used $\mu_{LMC}=18.477 \pm 0.004$ (stat.) $\pm 0.026$ (syst.) mag recently obtained from an analysis of detached eclipsing binary stars \citep{pietrzynski2019distance} instead of $\mu_{LMC}=18.50$ mag adopted by \cite{freedman2001final}.

%The overall systematic error in the distance modulus to NGC~4414 is $0.16$ mag \citep{gibson2000hubble,ferrarese2000hubble,freedman2001final}, and it is mostly dominated by the $0.13$ mag in the systematic error of the LMC distance modulus, despite the significantly smaller systematic error of $0.026$ recently obtained from an analysis of detached eclipsing binary stars \citep{pietrzynski2019distance}. A new analysis from the Cepheids PL relation in the Milky Way using the upcoming Gaia DR4 data will be required to reduce significantly the NGC~4414, see Sec. 4 for a discussion. We note that instead of the $\mu_{LMC}=18.50$ adopted by \cite{freedman2001final} we used 18.477 mag \citep{pietrzynski2019distance}.

Second, we used the same cosmological sample as \citetalias{khetan2021new}, composed of 96 SNe extracted from the Combined Pantheon Sample \citep{scolnic2018complete}. We did not apply an upper limit to the SN redshifts, but we selected those SNe from the subsample with good quality photometric data that allows us to adequately sample the LC \citepalias[see more details in][]{khetan2021new}. The final redshift range of the sample is 0.02 < z < 0.075. The lower $z$ cut was selected to minimize impact of peculiar velocities uncertainties. We performed a hyerarchical Bayesian regression to infer the model parameters from a likelihood function of the cosmological SN sample, and the two siblings SNe, acting as calibrator sample. Following the methodology by \citetalias{khetan2021new}, the posterior of the model parameters is given by Bayes' theorem as

\begin{equation}
\label{eqn:bayes}
P(\Theta|D)\propto P(D|\Theta)P(\Theta).
\end{equation}

In our study, the LC fit parameters $B_{max}$, $V_{max}$, and $\Delta m_{15}$ corresponding to the observable SN properties are included in the vector $D$, and the model parameters $P_{0}$, $P_{1}$, $R$, $H_{0}$ are represented by the vector $\Theta$. Their Eq. $6$ gives the likelihood probability distribution $P(D|\Theta)$ after marginalising over all $\Theta_{i}$, computed for each SN, that incorporated the two calibrators and the cosmological sample. In our case, $N_{calib} = 2$ and $\mu_{calib}=\mu_{NGC4414}$. We used their Eq.$(4)$ for the SNe cosmological sample to substitute the distance modulus in Eq.\ref{eqn:tripp} as a function of redshift and $H_{0}$. They considered a cosmological model with a Robertson-Walker metric and a spatially flat Universe \citep{weinberg1972gravitation,visser2005classical}. The values for cosmic deceleration and cosmic jerk are $q_{0}=-0.55$ and $j_{0}=1$, respectively \citep{2020A&A...641A...6P}. In order to derive $H_{0}$, we consider a hierarchical Bayesian approach consisting of two sub-models, one for the SNe cosmological sample and another for the two sibling SNe in NGC~4414. The variance for each of the two SNe is:
\begin{equation}
\label{eqn:variance}
\sigma^2_{calib,i}=\sigma^2_{B,i}+\sigma^2_{\mu_{NGC4414}}+(P_{1}\sigma_{\Delta m_{15}})^2+2R(\sigma^2_{B,i}+\sigma^2_{V,i})-2R\sigma^2_{B,i}
\end{equation}

\begin{figure}[!ht]
\centering
\includegraphics[width=\columnwidth]{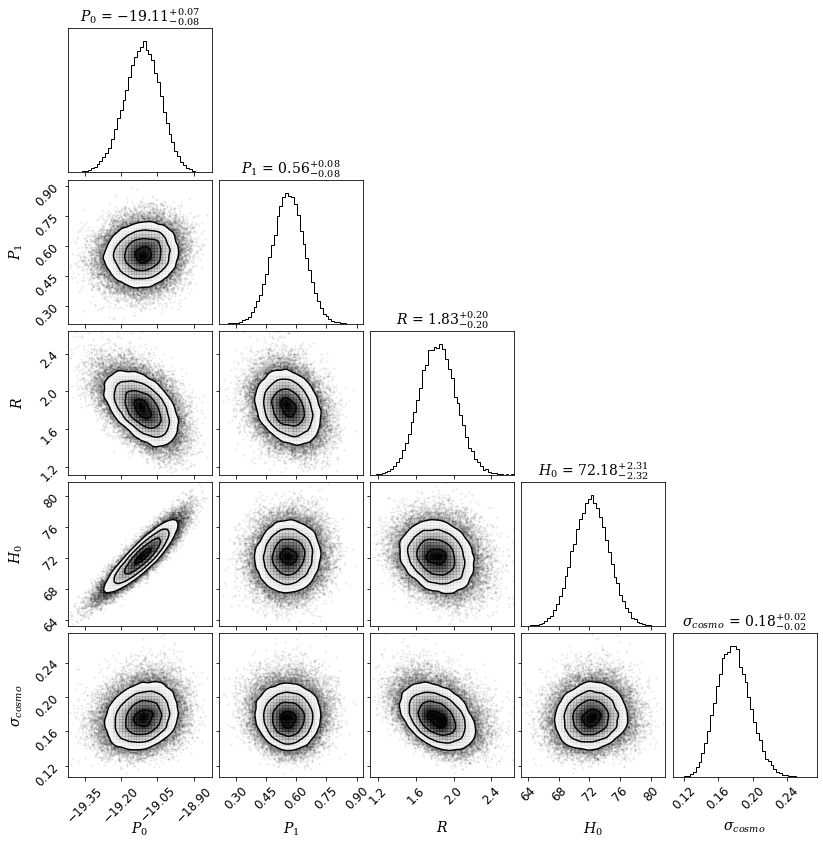}
\caption{\label{Fig:corner_plot} Corner plot showing the posterior probability distributions of the parameters $P_{0}$, $P_{1}$, $R$, $H_{0}$ and $\sigma_{int,cosmo}$. The median values of the posterior distributions are displayed on the histograms for each model parameter.} 
\end{figure}

We used Eq.$(11)$ in \citetalias{khetan2021new} to calculate the variance of each SN in the cosmological sample. This equation includes the term $\sigma_{int,cosmo}$ which accounts for any unknown scatter in the sample. For the priors, we considered a Half Cauchy distribution for $\sigma_{int,cosmo}$ and normal distributions for the rest of the parameters. We obtained the best-fit parameters of the model $P_{0}$, $P_{1}$, $R$, $H_{0}$, and $\sigma_{int,cosmo}$ by performing a Markov Chain Monte Carlo (MCMC) sampling implemented in Python with the PyMC3 probabilistic programming framework \citep{pymc}. Figure \ref{Fig:corner_plot} shows the posterior probability distribution (PPD) of the parameters inferred with the No-U-Turn Sampler (NUTS) algorithm\footnote{See more information in \url{https://docs.pymc.io/en/v3/}}. Table\,\ref{Tab:Modeling_param} shows the mean PPD values for the parameters using the two SNe as calibrators in our analysis, separately and together. 

We adopted the systematic uncertainties in $H_{0}$ described in \cite{freedman2001final}, but following the updated and improved analysis provided in \cite{freedman2012carnegie} (see more details in Appendix \ref{uncert}). We estimated the total systematic error in $H_{0}$ by adding in quadrature the systematic errors related to the distance modulus to NGC~4414 and those from the fits to the SNe light curves (see Table\,\ref{Tab:syst_errors}). After taking into account all the sources of uncertainties, the estimated value of $H_{0}$ from the two sibling SNe in NGC~4414 is $72.19 \pm 2.32$ (stat.) $\pm 3.42$ (syst.) km s$^{-1}$Mpc$^{-1}$. Although the systematic error is larger than the statistical error in our measurement, it can be reduced in the future, as we argue in Sec. 4. Table\,\ref{Tab:Modeling_param} also provides the Hubble constant estimates using each of the two SNe separately as calibrators, and compares with the $H_{0}$ for SN~1974 from the literature.

%\begin{figure*}[!htb]

\begin{figure} [ht!]
\includegraphics[width=\columnwidth]{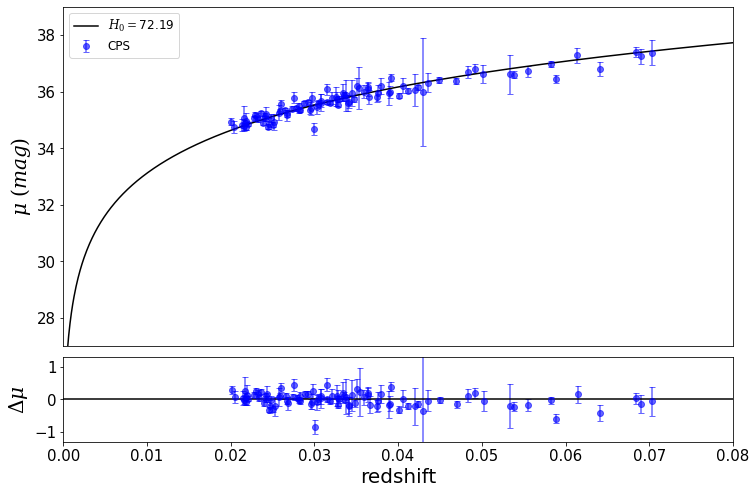}
\caption{Hubble diagram for the cosmological sample of $96$ SNe extracted from the Combined Pantheon Sample (CPS) and calibrated using the two sibling type-Ia SNe. The black line represents the resulting best-fit model. The lower panel shows the residuals from the model fit.}
  \label{Fig:hubble_diagram}
\end{figure}

\begin{table*}[!htb]
\centering
\caption{Posterior probability of the modeling parameters. The first two rows show the mean values obtained for each of the SNe separately in our analysis, and the third row considering both SNe. The last column shows the values of $H_{0}$ for SN~1974 derived from the literature.}
\label{Tab:Modeling_param} 
\begin{tabular}{lllllll}
\hline
\hline
Supernova & $P_{0}$  &  $P_{1}$  & $R$ & $\sigma_{cosmo}$ & $H_{0}$& $H_{0}(pub)$\\
 & (mag) &  (mag) &  &  & $(kms^{-1}Mpc^{-1})$& $(kms^{-1}Mpc^{-1})$\\
\hline
SN~1974G & $-19.17 \pm 0.10$ & $0.57 \pm 0.08$ & $1.81 \pm 0.20$ & $0.18 \pm 0.02$ & $70.19 \pm 3.32$ & $ 55 \pm 8$\tablefootmark{1}; $ 61.1 \pm 6.5$\tablefootmark{2}\\
SN~2021J & $-19.06 \pm 0.10$ & $0.57 \pm 0.08$ & $1.81 \pm 0.21$ & $0.18 \pm 0.02$ & $74.02 \pm 3.13$ & -\\
Both SNe & $-19.11 \pm 0.07$ & $0.56 \pm 0.08$ & $1.83 \pm 0.20$ & $0.18 \pm 0.02$ & $72.19 \pm 2.32$ &-\\

\hline
\end{tabular}
\tablefoot{
Value of the Hubble constant for SN~1974G from the literature.
\tablefoottext{1}{ \cite{schaefer1998peak}}
\tablefoottext{2}{ \cite{gibson2000hubble}}
}
\end{table*}

%-----------------------------------------------------------------

\section{Discussion and conclusions}
\label{conclusions}
In this study, we revisit the LC analysis of the known type-Ia SN~1974G in the nearby galaxy NGC~4414. This single calibrator yields a value of the Hubble constant $H_{0}=70.19 \pm 3.32$ (stat.) km s$^{-1}$Mpc$^{-1}$, as presented in Sec. 3.2. We obtain a more accurate and precise value of $H_{0}$, as compared to the significantly lower estimates reported in the literature (see Table\,\ref{Tab:Modeling_param}), by introducing important improvements in our analysis. First, we improved the methodology by applying the SNooPy LC fitter to determine more accurately the LC parameters for all SNe~Ia in the cosmological sample and the two sibling SNe in NGC~4414 (SN~1974G and the recently discovered SN~2021J). Secondly, we constructed the Hubble diagram using our sibling SNe (together and separately) as the zero-point calibrators. We used a distant SNe~Ia sample extracted from the Combined Pantheon Sample, in the redshift range $z = 0.02-0.075$, instead of using the Cal\'an-Tololo \citep{hamuy1996absolute} considered in those SN~1974G previous works. Our cosmological sample is of higher quality and has a larger number of SNe compared with the Cal\'an-Tololo sample ($96$ vs. $29$), which improves the statistical uncertainties in the $H_{0}$ estimates. 

In addition to SN~1974G, we used a recent supernova SN~2021J in the same galaxy as a calibrator. These "siblings" offer additional advantages by sharing several sources of systematics that in general have a non-negligible weight in the estimation of the distance, and then in the calibration of the luminosity relations of the SN. Among these, we have not only the same host distance, but also the peculiar velocity, and host galaxy properties that can have an influence on the luminosity, such as the extinction. However, while \cite{scolnic2021pantheon+} find correlations in some SN properties for SNe hosted in the same galaxy, such as the stretch parameter, they do not find greater correlation in the distance modulus values computed by siblings than in any other SN pairs.

Therefore, we have significantly reduced the statistical uncertainties, down to 3.2\%, in the determination of $H_{0}$ from the two sibling SNe in NGC~4414. However, our estimate of $H_0 = 72.19 \pm 2.32$ (stat.) $\pm 3.42$ (syst.) km s$^{-1}$Mpc$^{-1}$ still has large systematic errors in determining the distance modulus to NGC~4414, as explained in Sec. 3.2. We believe that a new Milky Way Cepheid PL calibration using the upcoming Gaia DR4, together with additional Cepheids from new HST observations, will significantly improve the systematic error of the NGC~4414 distance modulus, and hence provide an accurate determination of $H_{0}$ at the level of the statistical error.

We have also explored the possibility of measuring the distance modulus to NGC~4414 using the Tip of the Red Giant Branch (TRGB) method. Considering that there are no deep images sampling the stellar halo of the galaxy in the HST archive, it does not seem possible to estimate a distance with this method (Barry Madore, Wendy Freedman, private communications). NGC~4414 may be a good target for the upcoming James Webb space telescope (JWST).

%Figure\,\ref{Fig:hubble_diagram} shows the resulting Hubble diagram for the cosmological sample. The distance modules are calculated from the peak luminosity obtained with our Bayesian model calibrated with SN~1974G and SN~2021J. The redshift values are given in the CMB rest frame. The black line corresponds to the model for the best-fit parameters (see Table\,\ref{Tab:Modeling_param}).

%The relative error of $H_{0}$ has decreased significantly from the 14.5\% estimated by \cite{schaefer1998peak} to 8.2\%. If we consider the relative statistical error alone, it is reduced to 3.2\% by considering the sibling, and is comparable to the value of 3.4\% obtained using SBF \citepalias{khetan2021new}.

Our estimate of the Hubble constant from the two sibling SNe in NGC~4414 is consistent, within $\sim 1\sigma$ with that obtained from early-universe CMB measurements \citep{2020A&A...641A...6P} and local estimates from Cepheid variables \citep{riess2021comprehensive,brout2022pantheon+}, TRGBs \citep{freedman2021measurements}, and SBF \citepalias{khetan2021new}. Before requiring additional new physics beyond the standard cosmological model, it is imperative to further study possible systematic errors and provide more reliable measurements of $H_{0}$ using different methodologies in the local Universe. The possibility of finding at least three more pairs of sibling SNe~Ia in galaxies with Cepheid distances would not be negligible in the next decade. The ZTF survey \citep{Bellm2019} discoveries reinforce the previous premise, reporting 1-2 SNe~Ia every year in the 20 Mpc volume \citep{2022arXiv220304241D}, where HST observations can provide robust distance estimates by means of Cepheids or TRGB. As a back-of-the-envelope calculation, this means reducing the statistical error of $H_{0}$ down to 1.7\%, similar to what is expected from the entire sample of SNe~Ia, and more importantly, it would strongly reduce the systematic error, given that it mainly depends on the uncertainty on the distance modulus of a small number of host galaxies. We finally recall that the systematic error of $H_{0}$ using these sibling SNe also represents an upper limit since it will be greatly reduced using the incoming Gaia calibrations.

%\begin{figure}[!htb]
%\includegraphics[width=\columnwidth]{hubble_comparison_new.png}
%\caption{\label{Fig:hubble_comparison} Comparison of several values of the Hubble constant %obtained with different methods. We include recent values from the literature for late Universe, %either using Cepheids \citep{riess2021comprehensive,brout2022pantheon+}, SBF %\citepalias{khetan2021new}, or TRGB \citep{freedman2020calibration} as calibrators in galaxies %hosting SNe~Ia. The cyan dashed line represents the value of $H_{0}$ based on early Universe data %from observations of the CMB \citep{2020A&A...641A...6P}.}
%\end{figure}

\begin{acknowledgements}
We thank the anonymous referee for the helpful comments. We thank Barry Madore, Wendy Freedman, Radek Wojtak, Eoin \'O Colg\'ain, and Adam G. Riess for useful comments and discussions. The authors acknowledge financial support from the State Agency for Research of the Spanish MICINN through the "Center of Excellence Severo Ochoa" award to the Instituto de Astrof\'isica de Andaluc\'ia (SEV-2017-0709). EGC thanks the incentive grant by the Ministry of Economic Transformation, Industry, Knowledge and Universities REF. P20\_00880, project co-financed by 80\% by funds from the operational program FEDER de Andaluc\'ia 2014-2020. LI and NK were supported by two grants from VILLUM FONDEN (project number 16599 and 25501). CDT, FP and EP thank the support of the Spanish Ministry of Science and Innovation funding grant PGC2018-101931-B-I00. CDT thanks the Instituto de Astrof\'isica de Canarias for the use of its facilities to carry out this work. Based on observations collected at the Observatorio de Sierra Nevada, operated by the Instituto de Astrof\'isica de Andaluc\'ia (IAA-CSIC). Partly based on observations made with the Gran Telescopio Canarias (GTC), installed at the Spanish Observatorio del Roque de los Muchachos of the Instituto de Astrof\'isica de Canarias, in the island of La Palma. The authors thank the GTC support staff for performing the observations.
\end{acknowledgements}


\bibliographystyle{aa} % style aa.bst
\bibliography{bibliography.bib} % your references Yourfile.bib

\begin{appendix}

\section{Systematic uncertainties on the Hubble constant}
\label{uncert}
We adopted the systematic uncertainties in $H_{0}$ described in \cite{freedman2001final} and \cite{freedman2012carnegie}, who provided the Cepheid distance to NGC 4414 used in this study. The former detected Cepheid variables and derived accurate distances to nearby SN Ia hosts, including NGC 4414. The latter focused mainly on the zero point of the Cepheid extragalactic distance scale and a re-evaluation of the systematic error budget, and hence improving the systematic errors. 

We show in Table\,\ref{Tab:syst_errors} the systematic error budget on $H_{0}$ \citep[see more details in][]{freedman2012carnegie}. The systematics of the SNooPy fits were obtained in this work. The total error in $H_{0}$ was then calculated by combining all the individual errors in quadrature. The choice of the uncertainty in the metallicity parameter is controversial due to the value adopted for the Cepheid metallicity coefficient \citep[see Fig.14 in][]{gerke2011study}. Although there are works with smaller values \citep[e.g.][]{riess2021cosmic}, we conservatively considered the 4\% error adopted in \cite{freedman2001final}, since we used the distance to the galaxy from this work. We adopted a 1\% uncertainty due to crowding, as reported in \cite{freedman2012carnegie}. They present a quantitative analysis of the crowding bias in Cepheid photometry using artificial stars. The blue main sequence stars are the main contributors to the crowding, which make Cepheid photometry brighter and bluer. The systematic bias closely tracks the line of the Wesenhelt reddening free index (W) in the optical CMD, resulting in a minimal change in the final Cepheid distance (see more details in their Section 3.3 and their Appendix). \cite{freedman2012carnegie} conclude that the crowding effects are less than 0.02 mag (1\% in distance), in agreement with the results of \cite{2000PASP..112..177F}. Our estimated value of $H_{0}$ and its uncertainties from the two sibling SNe in NGC~4414 is $72.19 \pm 2.32$ (stat.) $\pm 3.42$ (syst.) km s$^{-1}$Mpc$^{-1}$.

\begin{table}[!htb]
\centering
\caption{Budget of systematic errors on $H_{0}$ from the 2 sibling SNe in NGC~4414}
\label{Tab:syst_errors} 
\begin{tabular}{lll}
\hline
\hline
Source of uncertainty & Error  &  Magnitude\\
 & \%  &  (mag)\\
\hline
LMC zero-point & $1.2$ & $0.026$\\
WFPC2 zero-point & $1$ & $0.022$ \\
Reddening & $1$ & $0.022$ \\
Metallicity & $4$ & $0.087$ \\
Bias in Cepheid PL & $1$ & $0.022$ \\
Crowding & $1$ & $0.022$ \\
B-band fit SNooPy & $0.55$ & $0.012$ \\
V-band fit SNooPy & $0.87$ & $0.019$\\
\hline
Total & $4.74$ & $0.103$ \\
\hline

\end{tabular}

\end{table}

\section{Photometry of SN~2021J}
In this section, we report the photometry of  SN~2021J. On the one hand, Table\,\ref{Tab:photometry} shows BVRI photometry of SN~2021J in eight epochs obtained with the CCDT150 camera mounted on the 1.5m telescope at the Observatorio de Sierra Nevada (OSN). On the other hand, Table\,\ref{Tab:swift} presents UBV photometry observed by {\it Swift}-UVOT.
\begin{table*}[!htb]
\centering
\caption{Photometry of SN~2021J obtained with the CCDT150 camera mounted on the 1.5 meter telescope at the OSN.}
\label{Tab:photometry} 
\begin{tabular}{lllll}
\hline
\hline
MJD & $B$  &  $V$  & $R$ & $I$\\
(days) & (mag) &  (mag) & (mag) & (mag)\\
\hline
$59227.17$ & $12.839 \pm 0.051$ & $12.653 \pm 0.041$ & $12.465 \pm 0.041$ & $12.516 \pm 0.068$\\
$59230.13$ &	$12.488 \pm 0.056$ & $12.278 \pm 0.079$ & $12.247 \pm 0.071$ & $12.247 \pm 0.071$\\
$59234.06$ &	$12.406 \pm 0.092$ & $12.426 \pm 0.060$ & $12.149 \pm 0.067$ & $12.280 \pm 0.083$\\
$59242.18$ &	$13.142 \pm 0.023$ & $12.367 \pm 0.033$ & $12.607 \pm 0.049$ & $12.733 \pm 0.081$\\
$59248.13$ &	$13.502 \pm 0.056$ & $12.702 \pm 0.050$ & $12.624 \pm 0.071$ & $12.527 \pm 0.081$\\
$59260.11$ &	$14.813 \pm 0.064$ & $13.627 \pm 0.042$ & $13.154 \pm 0.047$ & $12.696 \pm 0.079$\\
$59264.02$ &	$14.638 \pm 0.064$ & $13.650 \pm 0.042$ & $13.271 \pm 0.047$ & $12.804 \pm 0.079$\\
$59287.21$ &	$15.532 \pm 0.257$ & $14.612 \pm 0.128$ & $14.206 \pm 0.162$ & $13.809 \pm 0.197$\\

\hline
\end{tabular}

\end{table*}

\begin{table}[!htb]
\centering
\caption{Photometry of SN~2021J obtained by {\it Swift} Ultraviolet/Optical Telescope.}
\label{Tab:swift} 
\begin{tabular}{llll}
\hline
\hline
MJD & U\_UVOT  &  B\_UVOT  & V\_UVOT\\
(days) & (mag) &  (mag) & (mag)  \\
\hline
$59219.17$ & $15.35 \pm 0.08$ & $14.99  \pm 0.03$ & - \\
$59220.61$ & $14.74 \pm 0.06$ & $14.30  \pm 0.04$ & $13.61 \pm 0.05$ \\
$59221.80$ & $14.17 \pm 0.06$ & $13.45  \pm 0.03$ & $13.13 \pm 0.04$ \\
$59222.60$ & $13.95 \pm 0.06$ & $13.60  \pm 0.03$ & $12.94 \pm 0.03$ \\
$59223.66$ & $13.41 \pm 0.05$ & $13.13  \pm 0.02$ & $12.71 \pm 0.03$ \\
$59225.20$ & $13.02 \pm 0.03$ & $13.04  \pm 0.06$ & - \\
$59230.16$ & $12.49 \pm 0.03$ & $12.55  \pm 0.03$ & $12.22 \pm 0.03$ \\
$59230.56$ & $12.50 \pm 0.03$ & - & -\\
$59232.42$ & $12.62 \pm 0.03$ & $12.58  \pm 0.02$ & $12.24 \pm 0.03$ \\
$59234.61$ & $12.78 \pm 0.03$ & $12.62  \pm 0.02$ & $12.29 \pm  0.03$ \\
$59241.14$ & $13.51 \pm 0.06$ & - & - \\
$59245.25$ & $14.02 \pm 0.06$ & - & $12.81 \pm  0.04$ \\
$59249.03$ & $14.41 \pm  0.06$ & - & $13.16 \pm  0.03$ \\
$59253.22$ & $14.88 \pm 0.11$ & $13.95  \pm 0.04$ & - \\

\hline
\end{tabular}

\end{table}

\section{EBV\_model2 LC fitting for SN~2021J}
\label{EBV_model}
\begin{figure} [ht!]
\includegraphics[width=0.90\columnwidth]{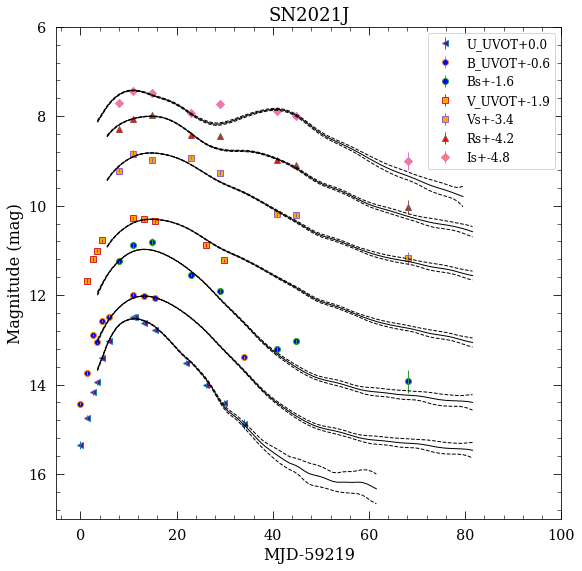}
\centering
\caption{\label{Fig:EBV_SN2021J} SN~2021J LCs fits using the EVB\_model in SNooPy. The legend indicates the color and shape of the symbols adopted for the different bands. The black lines indicate the SNooPy best-fits. Dashed lines are 1-$\sigma$ errors.}
  
\end{figure}

As seen previously, we used the \textit{max-model} in SNooPy to fit the LCs for the SNe, which allowed us to compute the maximum magnitude in each filter and $\bigtriangleup m_{15}$ needed for the Bayesian analysis. In this section, we calculate the LC parameters for SN~2021J by using another fitting model: the \textit{EBV\_model2}. This model allows us to compute the distance and the dust extinction of the host galaxy following the work of \cite{2006ApJ...647..501P}. We parameterized the SNe LCs using the \textit{color-stretch parameter} $s_{BV}$ introduced by \cite{2014ApJ...789...32B} and defined as the time between $B_{max}$ and ($B_{max}-V_{max}$) divided by $30$. We obtained a value of the reddening due to the host galaxy of $E(B-V)_{host}=0.349 \pm0.018$ mag, $s_{BV}=0.988 \pm0.024$, and $\mu_{NGC4414}=30.927 \pm0.053$ mag. Figure\,\ref{Fig:EBV_SN2021J} shows the LCs fits obtained. Fitting this model to just two filters for SN~1974G is not accurate to correctly disentangle the intrinsic extinction. Therefore, we decided to use the \textit{max-model} that meets the requirements of our work. 

\section{Uncertainty analysis of SN~1974G}
\label{unc_SN1974G}
\begin{figure*}[!ht]
\centering
\includegraphics[width=\textwidth]{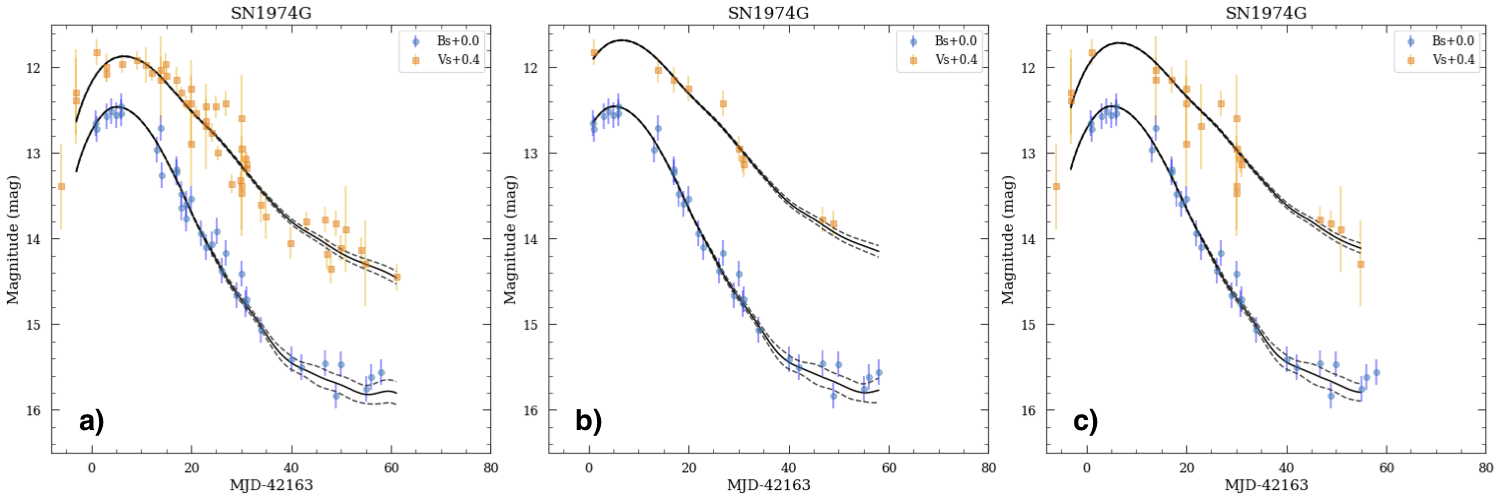}
\caption{\label{Fig:comparison_SN1974G} SN~1974G LCs fits using SNooPy. The black lines indicate the SNooPy best-fits. Dashed lines are 1-$\sigma$ errors. (a) Considering all the point used by \citet{schaefer1998peak}. (b) Eliminating the visual, photographic, and Burgat's points. (c) Eliminating the visual, and photographic points.}
\end{figure*}

In this section, we study the uncertainties corresponding to SN~1974G. As we seen in previous sections, we used the $B-$ and $V-$band photometry of SN~1974G presented in \citet{schaefer1998peak}. Due to the existing data contains both visual and photographic magnitudes, we studied our results in different cases, depending on what points are included in our analysis. Figure\,\ref{Fig:comparison_SN1974G} shows the LCs fits for SN~1974G for the different analysis. The solid black lines are SNooPy fits for the different filters. Dashed lines in LC fits indicate the 1-$\sigma$ errors. In a) we have considered all the points that \citet{schaefer1998peak} adopted. In b) we have removed the visual, photographic, and points with large uncertainty (Burgat's points, see Table 3 in \citet{schaefer1998peak}). In c) we have removed only the visual, and photographic points. Table\,\ref{Tab:analysis_SN1974G} shows the LC parameters for SN~1974G. The results are still consistent within the margins of error. Finally, we decided to consider all the original points because the SN peak is better covered. Furthermore, the large uncertainties of SN~1974G are within our luminosity calibration code and reflected in the different weights of the two SNe.

\begin{table*}[!htb]
\centering
\caption{SNooPy best-fit parameters for the LCs of SN~1974G. (a) Considering all the point used by \citet{schaefer1998peak}. (b) Eliminating the visual, photographic, and Burgat's points. (c) Eliminating the visual, and photographic points.}
\label{Tab:analysis_SN1974G} 
\begin{tabular}{llll}
\hline
\hline
Parameter & SN1974G (a)  &  SN1974G (b)  & SN1974G (c) \\
\hline
$B_{max}$ (mag) & $12.417 \pm 0.043 \pm 0.012$ (syst.) & $12.409 \pm 0.032 \pm 0.012$ (syst.) & $12.408 \pm 0.032 \pm 0.012$ (syst.) \\
$V_{max}$ (mag) & $12.252 \pm 0.035 \pm 0.019$ (syst.) & $12.074 \pm 0.049 \pm 0.019$ (syst.) & $12.103 \pm  0.047 \pm 0.019$ (syst.)\\
$T_{max}$ (MJD) & $42168.645 \pm 0.500 \pm 0.340$ (syst.) & $42168.682 \pm 0.378 \pm 0.340$ (syst.) & $42168.599 \pm 0.357 \pm 0.340$ (syst.) \\
$\bigtriangleup m_{15}$ (mag) & $1.236 \pm 0.074$ & $1.200 \pm 0.064$ & $1.195 \pm 0.061$ \\

\hline
\end{tabular}
\end{table*}

\end{appendix}

\end{document}